\documentclass[aps, prl, twocolumn, superscriptaddress, groupedaddress]{revtex4}
\pdfoutput=1

\usepackage{amsmath}
\usepackage{graphicx}
\usepackage{hyperref}
\DeclareMathOperator{\sech}{sech}

\begin{document}

\title{Interaction of high order solitons with external dispersive
  waves}

\author{I. Oreshnikov}
\affiliation{Department of Physics, St. Petersburg State University, 198504, Ulyanovskaya st. 1, Peterhof, St. Petersburg, Russian Federation}
\author{R. Driben}
\affiliation{ITMO University 197101, Kronverksky pr. 49, St. Petersburg, Russian Federation}
\affiliation{Department of Physics and CeOPP, University of Paderborn, Warburger Str. 100, D-33098 Paderborn, Germany}
%\author[2]{A. V. Yulin}

%\affil[*]{Corresponding author: driben@mail.uni-paderborn.de}

\date{\today}
% \ociscodes{(190.4370) Nonlinear optics, fibers; (190.5530) Pulse propagation and temporal solitons.}
% \doi{\url{http://dx.doi.org/10.1364/ao.XX.XXXXXX}}

\begin{abstract}
  We have presented theoretical and numerical studies on interactions
  of dispersive waves with second order solitons. Disperive wave with
  considerable intensity resonantly colliding with the second order
  solitons can lead to acceleration/decceleration of the later with subsequent
  central frequency shifts, but still well preserve the oscillating
  structure of the 2-solitons. The 2-soliton creates an effective
  periodical refractive index profile and thus reflected and
  transmitted dispersive waves generate new spectral bands with sharp
  peak structures. When resonant dispersive waves interact with two
  second order solitons bounding them, multiple scattering with
  multiple frequency conversion of the radiation occurs. Thus
  2-solitons and radiation trapped in between them can produce
  effective solitonic cavity with "flat" or "concave mirrors"
  depending on the intensity of the input.
\end{abstract}

\maketitle

%\section{Introduction}
Optical solitons attract attention of researches for decades due to
the profound interest of their fundamental physics and technological
applications \cite{book}. In addition to the fundamental solitons,
integrable models and physical media described by nearly-integrable
equations \cite{RMP} support high order $N$-solitons, with $N\geq 2$,
which are oscillating pulses periodically restoring their shape at
distances that are multiples of the fundamental soliton period
\cite{Satsu}. Experimental realization of 2- and 3-order optical
solitons dates back to 1983 \cite{Stolen}. Initial narrowing of
higher-order solitons was observed for soliton's orders up to $13$
\cite{Mollenauer}. Higher-order solitons were also demonstrated in the
cavity of a mode-locked dye laser \cite{Salin}. Strongly oscillating
higher-order solitons find natural applications for the pulse
compression \cite{Chan, Li} and frequency conversion \cite{Lee}, while
the breakup of high order solitons initiates the extremely important
process of the supercontinuum generation \cite{SC}.

Propagation of optical pulse close to zero dispersion wavelength makes
important consideration of high order dispersion terms and in
particular the third order dispersion (TOD) term. Second order
solitons (2-solitons) were found to be practically robust under the
influence of moderate TOD \cite{Wai}, preserving their periodically
oscillating nature. Very recently resonant radiation of 2-solitons was
investigated \cite{DribenYulin} revealing fascinating structure of
radiation band consisting of multiple frequency peaks. While emitting
the mentioned above radiation 2-soliton remained its fundamental
features for many periods. Interestingly, oscillating dissipative
solitons featuring similar radiation behavior were reported too
\cite{KudlinskiA, KudlinskiB}.

In is relevant to mention that extensive studies were performed
considering scattering of weak DWs on fundamental solitons, with
soliton’s properties remaining invariant \cite{yulin, pre, Efimov,
Efimov2, Skryabinoverview, Conforti}. On the other hand, it was also
demonstrated that strong resonant collision of DWs with solitons can
accelerate or decelerate solitons. These accelerations manifest
themselves in frequency domain by upshifts or downshifts of central
frequencies of the solitons \cite{DribenMitschke, Demircan, Demircan2,
Tartara}. Generation of new bands of frequencies due to the scattering
of DWs from solitons can be viewed as an alternative technique for
generation of broad and coherent supercontinuum \cite{Demircan2}
without the high order soliton fission.

Taking one step further a very interesting scenario with DWs
interacting with two co-propagating solitons \cite{Resonator,
  Resonator2, Kudlinski}. DWs temporally located in between the
solitons interact with one soliton after the other bouncing of them.
This process can repeat itself many times resulting in creation of
solitonic cavity behaving like two mirrors for the DWs. In case of DWs
with considerable intensity, the two solitons get attracted one
towards the other, resulting in collision or even fusion
\cite{Fusion}. This phenomenon was proven to be responsible for the
appearance of multiple soliton knot patterns \cite{DribenMitschke,
  Resonator2, NC} during the complex supercontinuum generation
process. Such a solitonic cavity was very recently successfully
realized in experimental conditions \cite{Kudlinski}. Also a "convex
mirror" cavity with mutually repulsing via DW two dark solitons was
demonstrated \cite{darksolitons}.

In the present work we aim to study interactions of 2-solitons with
external dispersive waves (DWs), demonstrating a possibility to
manipulate propagation trajectories of the 2-solitons, to create broad
continuum with rich spectral properties and also to create solitonic
cavities consisting of two 2-solitons and DWs bouncing between them.

%\section{Interaction of 2-solitons with weak dispersive waves}

Dynamics of optical pulse in the vicinity of zero dispersion
wavelength (ZDW) is governed by nonlinear Schr\"odinger equation with
the inclusion of TOD.
\begin{equation}
\label{eq:NLS}
i \partial_{z} u
  - \frac{1}{2} \beta_{2} \partial^{2}_{t} u
  - \frac{i}{6} \beta_{3} \partial^{3}_{t}
  + \gamma |u|^2 u = 0
\end{equation}
where $\beta_{2}$ and $\beta_{3}$ designate second and third order
dispersion terms, while $\gamma$ represents the nonlinear parameter.
Considering standard fiber with $\lambda_{ZDW} = 1311~\text{nm}$ and
launching a soliton at central wavelength of $\lambda_{sol} =
1470~\text{nm}$, we will work with the following values of dispersion
parameters: $\beta_{2} = -14.2~\text{ps}^{2}/\text{km}$ and $\beta_{3}
= 0.087~\text{ps}^{3}/\text{km}$. The nonlinear coefficient is taken
$\gamma=2~\text{W}^{-1}\text{km}^{-1}$.

The injected light consist of two pulses launched with different
frequencies namely soliton and DW
\begin{equation*}
  u_{0}(t)=u_{sol}(t)+u_{DW}(t)
\end{equation*}
The 2-soliton has the initial form
\begin{equation*}
  u_{sol}(t) = 2 \sqrt{P_{0}} \sech(t/T_{0})
\end{equation*}
with temporal width of $T_{0}=62.5~\text{fs}$ and the peak power of
corresponding fundamental soliton $P_{0}=1817~\text{W}$. DW is given
by
\begin{equation*}
  u_{DW}(t)=
    A_{DW} \sech((t-t_{1}) / T_{1})
           \exp(- i \delta\omega (t - t_{1}))
\end{equation*}
Here $A_{DW}$ represent the amplitude of DW, $t_{1}$ the temporal
delay of the DW with the respect to the soliton, $T_{1}$ is the
temporal width of dispersive wave, and $\delta \omega$ is the angular
frequency deviation of the DW from the soliton given by $\delta
\omega= 2 \pi c (\lambda_{inc}^{-1}-\lambda_{sol}^{-1})$.

\begin{figure}[!htbp]
  \centering
  \includegraphics[width=8.0cm]{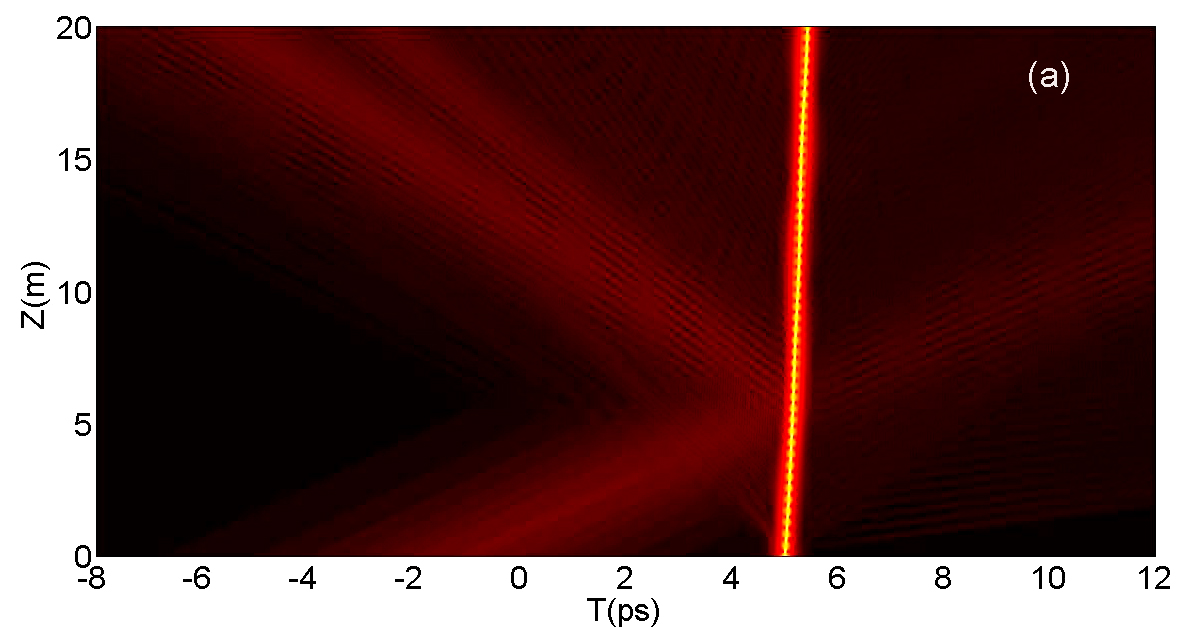}
  \includegraphics[width=8.0cm]{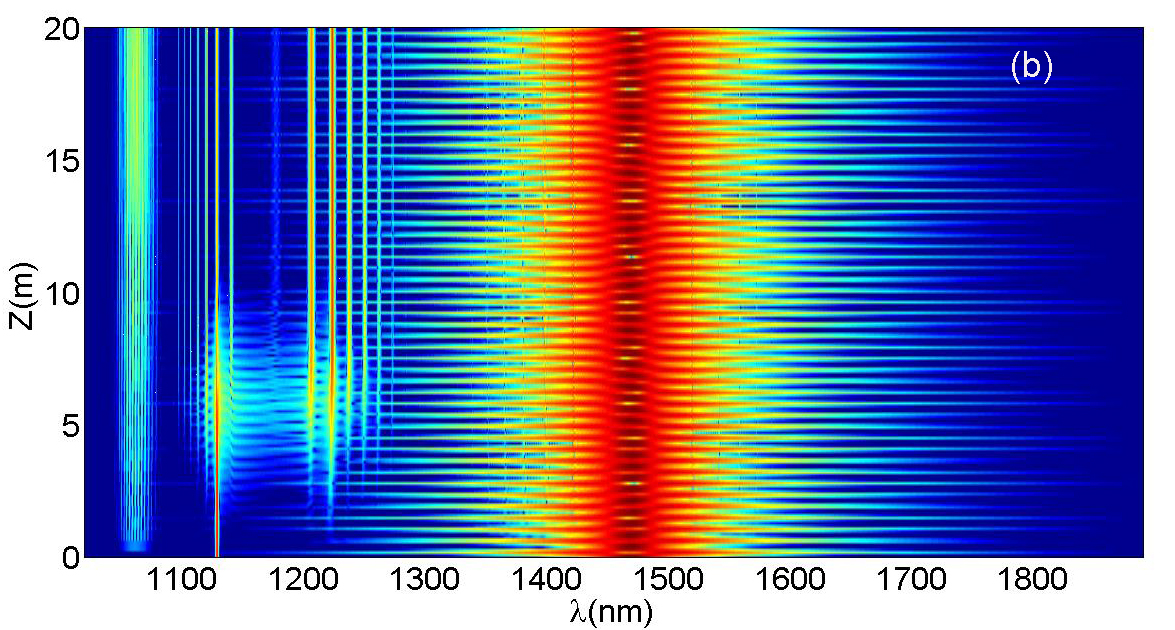}
  \includegraphics[width=8.0cm]{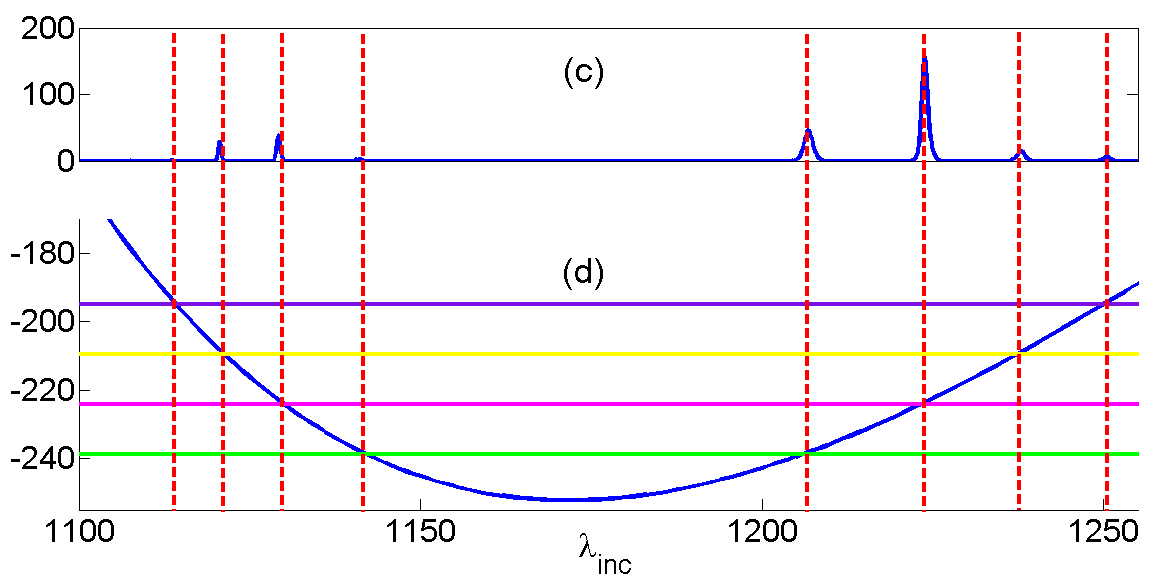}
  \caption{(Color online) Scattering of a wide DW on 2-soliton.
    Dynamics in (a) temporal domain with $|u|^{0.5}$ shown instead of
    the conventional intensity in order to observe better less intense
    regions. (b) Spectral domain dynamics presented in logarithmic
    scale. (c) The spectral density at the end output of the fiber
    with the region of reflected and transmitted DW zoomed (from
    $1100~\text{nm}$ to $1255~\text{nm}$). (d) is the graphical
    solution to \eqref{eq:ResonanceFrequencies} predicting the
    position of the resonance frequencies. Blue curve is the left hand
    side of the equation; horizontal lines correspond to the right
    hand's side for $N = -1,0, .. ,3$ starting from the bottom.}
  \label{ScatteringWideDW}
\end{figure}

The result of collision of a relatively wide DW ($T_{1} = 20 T_{0}$)
centered at $1130~\text{nm}$ and having a very small amplitude of
$A_{DW}^{2}=4.54~\text{W}$ with the 2-soliton is presented in
Fig. \ref{ScatteringWideDW}(a, b). Fig. \ref{ScatteringWideDW}(a)
demonstrates dynamics in temporal domain and it reveals that the DW is
partially scattered and partially transmitted through the 2-soliton
that creates an effective periodic refractive index potential. In
general one can archive much higher percentage of scattering by
optimizing the incident wavelength of the DW within the resonance
band, but here we aim to inspect also the transmitted part of the DW.
Looking at the spectral domain representation of the dynamics in Fig.
\ref{ScatteringWideDW}(b) we can see several new frequency bands
arising in a course of evolution. The leftmost band below
$1100~\text{nm}$ consisting of multiple peaks is the resonant or
Cherenkov radiation \cite{Akh} emitted form the 2-soliton. Its central
spectral position as well as the separation between the frequency
peaks is described by \cite{DribenYulin}:
\begin{equation*}
  \label{eq:CherenkovResonances}
  \frac{1}{2} \beta_{2} \delta\omega^{2}
    + \frac{1}{6} \beta_{3} \delta\omega^{3}
    - v_{g}^{-1} \delta\omega
    = \frac{1}{2} \gamma P
    + \frac{2 \pi}{Z_{0}} N
\end{equation*}
Here $\delta\omega = 2 \pi c (\lambda^{-1} - \lambda_{sol}^{-1})$ is
frequency detuning of the resonant wave from the soliton frequency,
$v_g$ is the soliton velocity, $Z_{0}$ is a period of the oscillations
of 2-soliton and N is an integer number.

The region of transmitted and reflected radiation at the output of the
fiber is shown in Fig. \ref{ScatteringWideDW}(c). The resonance
frequencies for the radiation generated by FWM mixing between incident
DW and 2-soliton were derived by using the perturbation theory similar
to the analysis for the fundamental soliton in \cite{pre}. In contrast
with the fundamental soliton case, the resonant radiation becomes
polychromatic, with frequency detunings for incident and scattered
fields $\delta\omega_{inc, ~sc} = 2 \pi c (\lambda_{inc, ~sc}^{-1} -
\lambda_{sol}^{-1})$ being tied by the following relation
\begin{equation}
  \label{eq:ResonanceFrequencies}
  \frac{1}{2} \beta_{2} \delta\omega_{sc}^{2}
    + \frac{1}{6} \beta_{3} \delta\omega_{sc}^{3}
    = \frac{1}{2} \beta_{2} \delta\omega_{inc}^{2}
    + \frac{1}{6} \beta_{3} \delta\omega_{inc}^{3}
    + \frac{2 \pi}{Z_{0}} N
\end{equation}
We solved the \eqref{eq:ResonanceFrequencies} graphically in order to
obtain the spectral positions of the transmitted and reflected peaks
and presented it in Fig. \ref{ScatteringWideDW}(d). For example the
intersection of the blue curve with the horizontal magenta line
represents the scattered and transmitted DWs of the zero order $(N =
0)$, that corresponds to the strongest reflected peak at about
$1223~\text{nm}$ and strongest transmitted peak at $1130~\text{nm}$.
The intersection of the blue curve with the horizontal green line
stand for $N = -1$ and so on. Comparison between these results and
those obtained by direct numerical simulations (Fig.
\ref{ScatteringWideDW}(c)) is visualized by vertical dashed lines. We
can find an excellent agreement between the two approaches with a
discrepancy of only few nanometers.

\begin{figure}[!htbp]
  \centering
  \includegraphics[width=8.8cm]{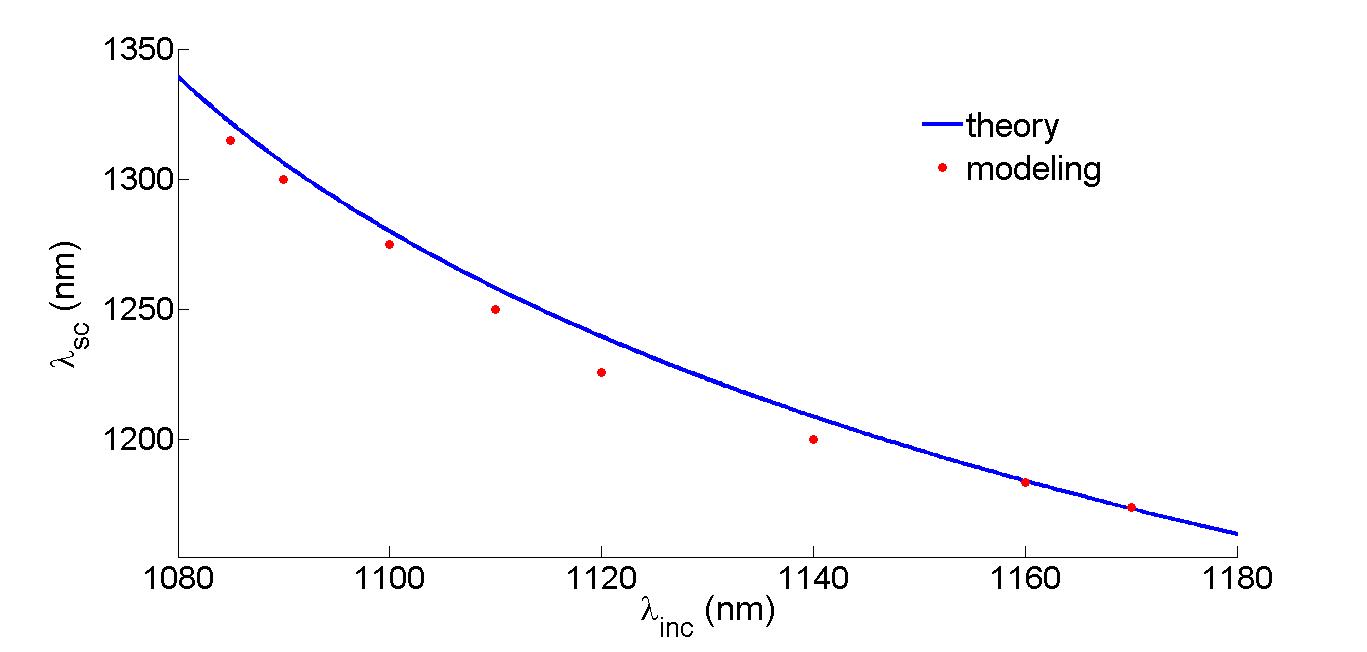}
  \caption{(Color online) Central wavelength of the scattered
    radiation as a function of incident. Solid line is the curve
    predicted by \eqref{eq:ResonanceFrequencies} for fundamental ($N =
    0$) harmonic. Square markers are found by numerical modeling.}
  \label{ScatteredVsIncident}
\end{figure}

Using more narrow incident DWs with the temporal width equal to that
of the soliton ($T1 = T0$) we have produced systematic simulations to
provide the dependence of the scattered radiation central wavelength
vs. the incident central wavelength as shown in Fig.
\ref{ScatteredVsIncident}. Solid curve is the plot of the analytical
relation $\lambda_{sc}(\lambda_{inc})$ that is obtained from graphical solution of
\eqref{eq:ResonanceFrequencies} for the fundamental harmonic $N = 0$.
Shifting $t_{1}$ and thus modifying the place of collision with the
respect to soliton's shape at the moment of the collision did not
significantly influence the results.

%\section{Interaction of strong dispersive waves with 2-solitons.%Generation of broadband light and creation of 2-soliton cavities}
Now we will inspect what happens when DW with more considerable
intensity collides with the 2-soliton. In this case the velocity of
the soliton will be modified and we can no longer rely on analytical
approximations, but to study only results of numerical simulations. We
collide DW with the 2-soliton as in the previous chapter, but this
time we increase its input intensity to $A_{DW}^{2}= 454.4 ~
\text{W}$, that is still several times smaller that the input peak
intensity of the 2-soliton participating in the collision and we will
take the width of the DW equal to that of the soliton ($T_{1} =
T_{0}$). A typical result of collision of strong DW with the 2-soliton
still resisting to the splitting is presented in the Fig.~3. The
soliton's trajectory is considerably shifted, consequently it central
wavelength is downshifted from $1470~\text{nm}$ to $1463~\text{nm}$.
We can also observe in Fig.~3 (lower panel) generation of significant
new light spectral bands in region between the soliton's central
wavelength and DW. Increasing the intensity of the incident DW much
higher we will be observing the splitting of the 2-soliton.

\begin{figure}[!htbp]
  \includegraphics[width=8.8cm]{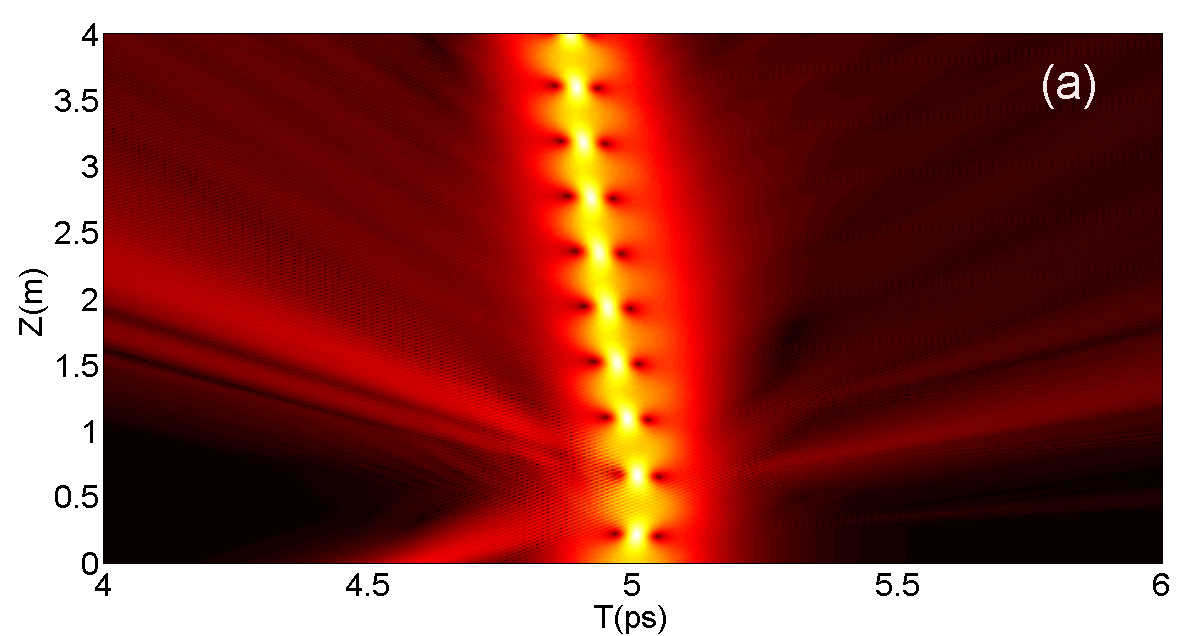}
  \includegraphics[width=8.8cm]{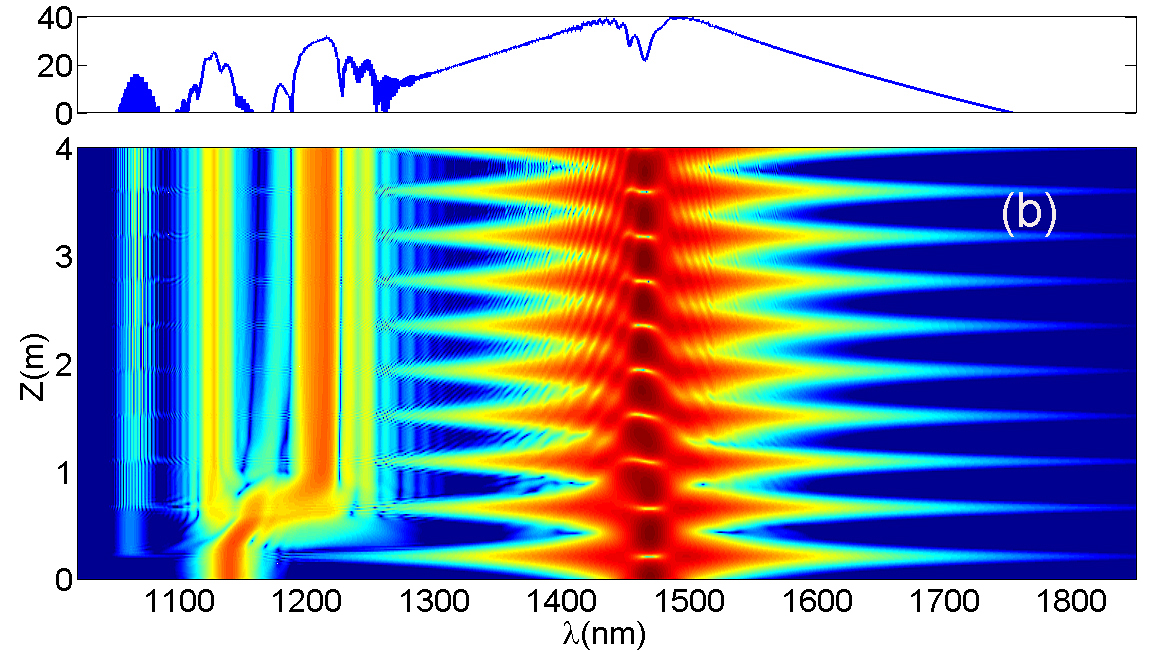}
  \caption{(Color online) Interaction of a strong DW with 2-soliton.
    Dynamics in (a) temporal domain showing $|u|^{0.5}$. (b) Spectral
    domain dynamics representation in logarithmic scale.}
  \label{fig3}
\end{figure}

\begin{figure}[!htbp]
  \centering
  \includegraphics[width=8.0cm]{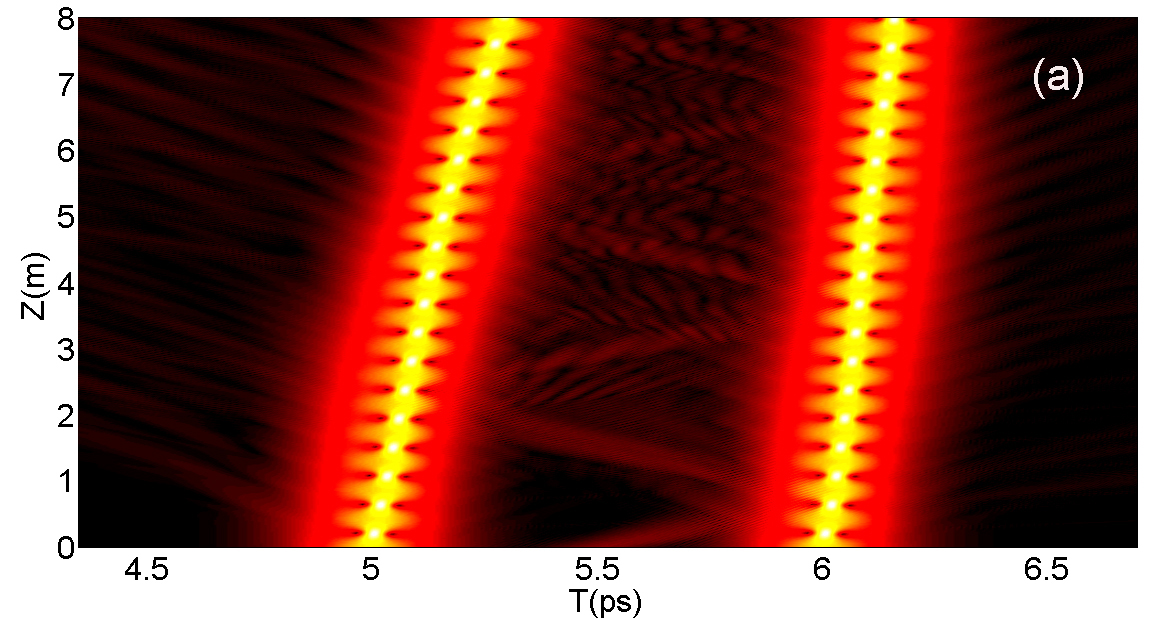}
  \includegraphics[width=8.0cm]{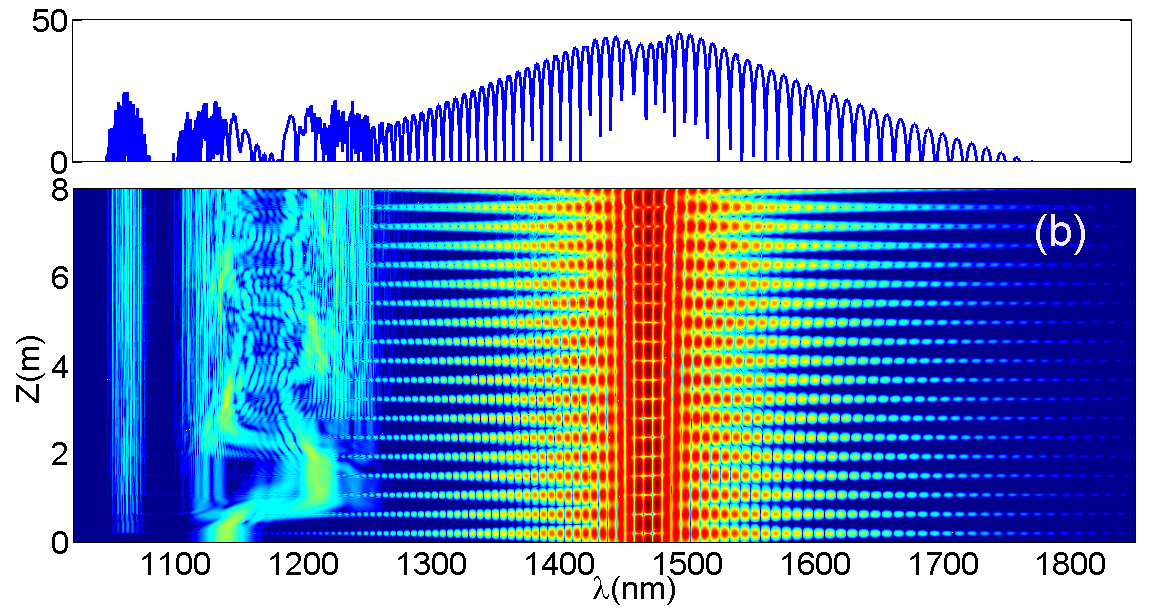}
  \caption{(Color online) Solitonic cavity created by two 2-solitons
    and DW bouncing between them. (a) Temporal domain representation
    with $|u|^{0.5}$ shown. (b) Spectral domain representation with
    output spectrum profile shown above the panel.}
  \label{CavityWeakDW}
\end{figure}

Even more interesting interaction scenario occurs when DW bounces off
two co-propagating 2-solitons. We consider an input consisting of two
well separated 2-solitons and DW launched in between them such as
\begin{align*}
  u_{0}(t)
    &= 2 \sqrt{P_{0}} \sech((t + t_{1}) / T_{0}) \\
    &+ A_{DW} \sech(t / T_{0}) \exp(- i \delta \omega t) \\
    &+ 2 \sqrt{P_{0}} \sech((t - t_{1}) / T_{0})
\end{align*}
If the wavelength of DW is located well inside the resonance band we
will be able to observe its multiple reflections from the 2-solitons
If the intensity of the DW is not very high the trajectories and the
spectral characteristics of the two solitons are almost not modified
and reflections occur pretty periodically to certain wavelength
regions. For example the DW initially launched at $1140~\text{nm}$
with $A_{DW}^{2}=18.17 ~\text{W}$ is being reflected several times to
the region around $1200~\text{nm}$ and then back to its original
spectral location (Fig. \ref{CavityWeakDW}(a, b)). Interestingly the
spectrum of the two solitons manifests multiple periodically
alternating light and dark regions spanned for this particular example
from $1250~\text{nm}$ to $1750~\text{nm}$. Such an illumination
spectrum that can be further controlled by the parameters of the
participating solitons can be suggested for photonics applications
requiring particular spectral design.

\begin{figure}[!htbp]
  \centering
  \includegraphics[width=8.8cm]{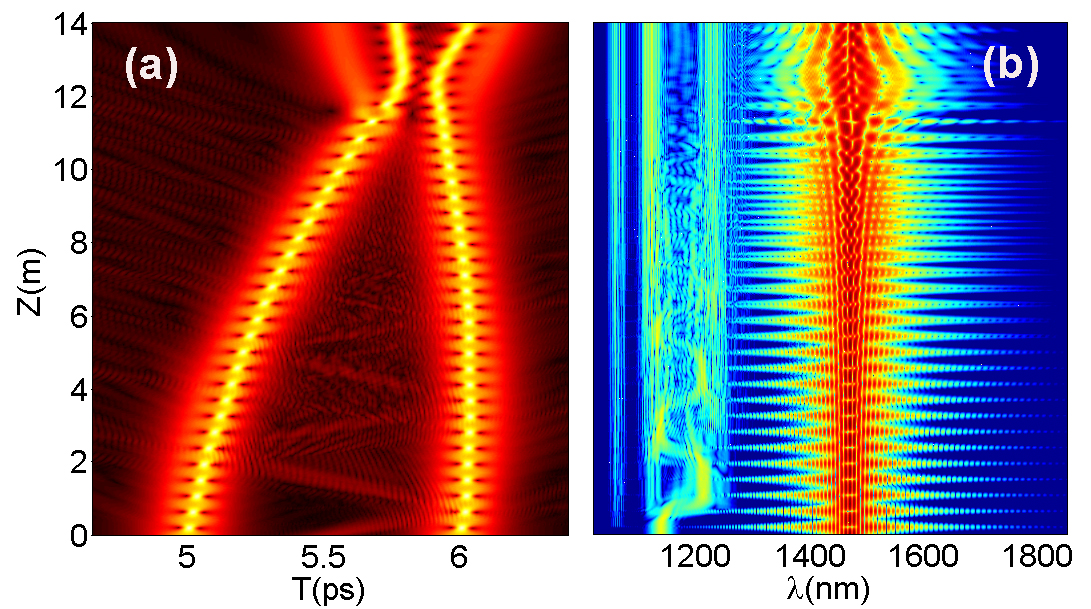}
  \caption{(Color online) Solitonic "concave mirror" cavity created by
    two 2-solitons and stronger DW bouncing between them. (a) Temporal
    domain representation with $|u|^{0.5}$ shown. (b) Spectral domain
    representation.}
  \label{CavityStrongDW}
\end{figure}

As mentioned above 2-solitons feature high degree of robustness when
colliding with DWs and thus we can build a closed cavity consisting of
two 2-solitons acting like concave mirrors without observing their
splitting. An example of such cavity is demonstrated in
Fig. \ref{CavityStrongDW}(a, b). Here we used the same type of input
light but increased an intensity of the incident DW to
$A_{DW}^{2}=90.85~\text{W}$ leaving the rest of the parameters
unchanged. In this example the central wavelength of the left and
right 2-solitons get upshifted and downshifted respectively. Also the
periodicity of illuminated regions such as in Fig. \ref{CavityWeakDW}
gets strongly distorted (Fig. \ref{CavityStrongDW}(b)).

To conclude, scattering of DWs on high order solitons was presented.
The structure of the 2-soliton provides an effective periodical
potential leading to polychromatic scattering and transmission for the
incident DWs. In spectral domain it manifest itself by the generation
of frequency comb like novel bands. Position of the peaks of the
transmitted and reflected bands is predicted by excellently
overlapping analytical and numerical approaches. Interaction of DWs
with two 2-solitons bounding them and creation of solitonic cavities
was also demonstrated. The presented nonlinear dynamics besides the
fundamental interest suggest prospective photonic devices schemes for
frequency conversion and broadband light generation with pre-designed
spectrum.

\section*{Funding Information}

R.D. gratefully acknowledges the support by the Russian Federation
Grant 074-U01 through ITMO Early Career Fellowship scheme.

\end{document}